\documentstyle{ws-art}
\begin{document}
\begin {center}
\noindent
Preprint TAUP-2176-94,~Sept. 1994,
Bulletin Board hep-ph@xxx.lanl.gov - 9409307\\
Contribution: Proceedings of the Conference
on Physics with GeV-Particle Beams,\\ Juelich, Germany, Aug. 1994,
World Scientific, Eds. H. Machner and K. Sistemich.\\[.3CM]
CHIRAL ANOMALY TESTS\\[.3CM]
Murray A. Moinester \\[.3CM]
\small{
\it{School of Physics and Astronomy,}\\
\it{Raymond and Beverly Sackler Faculty of Exact Sciences,}\\
\it{Tel Aviv University, 69978 Ramat Aviv, Israel}\\
\it{E-mail: murray@tauphy.tau.ac.il}}\\[1cm]
\end{center}
\begin{abstract}
\noindent
For the $\gamma$-$\pi$ interaction, the perturbative expansion of the
effective chiral Lagrangian ($\chi$PT) can be limited to terms quartic in
momenta and masses (O(p$^4$)), or to higher order. The abnormal intrinsic
parity (chiral anomaly) component of the lagrangian leads to interesting
predictions for the processes $\pi^0 \rightarrow  2 \gamma$ and $\gamma
\rightarrow  3 \pi$. These are
described by the amplitudes F$_{\pi}$ and F$_{3\pi}$,
respectively. We demonstrate that the O(p$^4$) value of F$_{3\pi}$ disagrees
with existing data, while the O(p$^6$) value is nearly consistent.
We describe how Fermilab
experiment E781 can get improved data for tests of the chiral anomaly.
\end{abstract}

The Chiral Axial Anomaly can be studied with a 600 GeV pion beam in FNAL
experiment E781 \cite{russ}.
For the $\gamma$-$\pi$ interaction, the O(p$^4$)
chiral lagrangian \cite {gl,donn1} includes Wess-Zumino-Witten (WZW) terms
\cite{wzw,bij3}, which lead to a chiral anomaly term \cite{wzw,bij3,anti}
in the
divergence equations of the currents. This leads directly to interesting
predictions \cite{bij3,ter} for the processes $\pi^0 \rightarrow  2 \gamma$
and $\gamma \rightarrow  3 \pi$; and other processes as well \cite{bij3,ben}.
The two
processes listed are described by the amplitudes F$_{\pi}$ and F$_{3\pi}$,
respectively. The F$_{\pi}$ vertex was first described by Adler, Bell, and
Jackiw \cite{abj}.

The chiral anomaly term leads to a prediction for F$_{\pi}$ and F$_{3\pi}$ in
terms of $N_c$, the number of colors in QCD; and f, the charged pion decay
constant. We use $N_c=3$ and f= 92.4 $\pm$ 0.2 MeV in
the equations given previously \cite {bij3,anti} for these amplitudes.
This f
value is from Holstein \cite{hols1}, and Marciano and Sirlin \cite {mar};
since the PDG  \cite{pdg}  value 93.2 does not account completely for
radiative corrections \cite {hols1,mar}.
The value we use differs from
f= 90. $\pm$ 5. MeV estimated by Antipov et al. \cite {anti},
and leads therefore also to different conclusions regarding the agreement of
data and theory.
The O(p$^4$) F$_{\pi}$ prediction \cite {bij3} is:
\begin{eqnarray}
F_{\pi} =  {\alpha N_c \over 3 \pi f} = 0.025~GeV^{-1},
\end{eqnarray}
in agreement with experiment \cite {bij3}. The F$_{3\pi}$ prediction
is:
\begin{eqnarray}
F_{3\pi} = {N_c (4 \pi \alpha)^{1 \over 2}
\over 12 \pi^2 f^3} \sim 9.7 \pm 0.2 ~GeV^{-3},~ O(p^4).
\end{eqnarray}
We estimate a theoretical
uncertainty of 0.2~GeV$^{-3}$ from f and including the accuracy of
the O(p$^4$) prediction. The latter is of
order \cite {gl,donn1}  m$_{\pi}^2/\Lambda^2$
$\sim 2\%$, where $\Lambda \sim 1 ~GeV$ sets the scale \cite {gl,donn1}
for the $\chi$PT
expansion. The O(p$^4$) relationship between these two amplitudes
was first given by Terentev \cite {ter}:
\begin{eqnarray}
F_{3\pi} = {F_{\pi} \over f^2 (4 \pi \alpha)^{1 \over 2}}.
\end{eqnarray}
The experimental confirmation of eq. 2 would demonstrate that the O(p$^4$)
terms are sufficient to describe F$_{3\pi}$.

   The amplitude F$_{3\pi}$ was measured
by Antipov et al. \cite{anti} at
Serpukhov with 40 GeV pions.
Their study involved
pion production by a pion in the nuclear Coulomb field via the
Primakoff
reaction:
$$
{\pi^-} + Z \rightarrow {\pi^-}'  + {\pi^0} + Z',  \eqno(4)
$$
where Z is the nuclear charge. The 4-momentum
of each particle is $P_{\pi}$, $P_Z$, $P_{{\pi}'}$, $P_{Z'}$, $P_{\pi^0}$,
respectively. In the one-photon exchange domain, eq. 4 is equivalent to:
$$
 {\pi^-} + \gamma  \rightarrow  {\pi^-}' + {\pi^0}, \eqno(5)
$$
and the 4-momentum of the virtual photon is k = $P_Z$-$P_{Z'}$. The
cross section formula for the eq. 4 reaction
was given in Ref. 6, and
depends on $F_{3\pi}^2$, and
on t, s, t$_1$, t$_0$, Z$^2$.
Here t is the squared four-momentum transfer to the
nucleus, $\sqrt{s}$ is the invariant mass of the $\pi^- \pi^0$
final state, t$_1$
is the squared 4-momentum transfer between initial and final $\pi^-$ in eq. 5,
$t_0$ is the minimum value of t to produce a mass $\sqrt{s}$,
and the virtual photon target density is proportional
to Z$^2$. The Antipov
et al. data sample (roughly 200 events)
covered the ranges $-t < 2. \times 10^{-3} (GeV/c)^2$ and
$s < 10.~m_{\pi}^2$. The small t-range selects
events predominantly associated with the exchange of a virtual photon,
for which the target nucleus acts as a spectator.

           The experiment \cite{anti} yielded
F$_{3\pi}=12.9 \pm
0.9 (stat) \pm 0.5 (sys) ~GeV^{-3}$. The uncertainties do not include
estimated 10\% errors \cite {anti,amen} arising from extrapolating F$_{3\pi}$
to threshold (s, t$_1$ approaching zero); for data taken in the s-range of
Antipov et al. The cited
experimental  result differs from the O(p$^4$) expectation
(eq. 2) by at least two standard deviations. Therefore, in contrast to the
conclusion of Antipov et al, we conclude
that the chiral anomaly prediction  at O(p$^4$) is
not confirmed by the available $\gamma \rightarrow 3\pi$ data.

Bijnens et al. \cite{bij3,bij1} studied higher order $\chi$PT
corrections in the abnormal intrinsic parity (anomalous) sector. They included
one-loop diagrams involving one vertex from the WZW term, and tree diagrams
from the O(p$^6$) lagrangian. They determine parameters of the lagrangian via
vector meson
dominance (VMD) calculations. The higher order corrections are small
for F$_{\pi}$. For F$_{3\pi}$, they increase the lowest order value
from 7\% to 12\%. The one-loop and  O(p$^6$) corrections to
F$_{3\pi}$ are comparable in strength. The loop corrections to F$_{3\pi}$ are
not constant over the whole phase space, due to dependences on the momenta of
the 3 pions. The average effect is roughly 10\%, which then changes the
theoretical prediction by 1. GeV$^{-3}$. Given the VMD assumption, we make a
rough uncertainty estimate of 30\% for this contribution. The
prediction,
including the errors given previously in eq. 2,
is then:
$$
F_{3\pi} \sim  10.7 \pm 0.5 ~ GeV^{-3}, ~O(p^6); \eqno(6)
$$
almost consistent
with the data. The limited accuracy of the
existing data, together with the new
calculations of Bijnens et al., motivate an improved and more precise
experiment.

We use the Primakoff cross section formula \cite{anti} for the reaction of eq.
(4), with the O(p$^6$) $F_{3\pi}$ value,  to calculate the expected cross
section for an incident 600 GeV energy. The cross section is about 100 nb for a
C$^{12}$ target for an s interval of 4-10 m$_{\pi}^2$; while the total
inelastic cross section is roughly 192 mb. The number of pion interactions in
the target during the E781 beam time is estimated to be about $3 \times
10^{10}$. Therefore, the expected number of two-pion
events for this s-interval is about $2. \times 10^4$.
The large number of events will allow analysis of the data separately in
different intervals of s. This is important because uncertainties \cite
{anti,amen} due to $\rho$ and $\omega$
contributions increase with s; and to control
systematic uncertainties.
The $\rho$ contributions (near the $\rho$-resonance s-value, and
near the two-pion threshold)
can be seen
in the data of Jensen et al. \cite {jens} for the Primakoff
reaction  ${\pi^-} + \gamma  \rightarrow  \rho^-
\rightarrow {\pi^-} + {\pi^0}$.
For reaction (4) in the s interval from 4-6 m$_{\pi}^2$, we expect
roughly an order of magnitude less events than in the
interval to 10 m$_{\pi}^2$.
This
number of events is still large enough to give excellent
statistical error.

Another reaction \cite{amen} to determine F$_{3\pi}$
also uses a virtual photon:
$$
\pi^- + e \rightarrow {\pi^-}' + \pi^0 +e',  \eqno(7)
$$
whereby an incident high energy pion scatters inelasticly from a target
electron in an atomic orbit. The number of such events
observed by Amendolia et al. \cite {amen} was 36 for a Hydrogen
target, corresponding to a cross section of 2.1 $\pm$
0.5~nb. The experiment did not extract a value for F$_{3\pi}$; but smaller
cross section uncertainties are in any case needed for a precision
chiral anomaly
test.
For this reaction on a carbon target,
as in Fermilab E781, the
expected cross section per atom
is roughly 10. nb., and the number of expected events
is
roughly 2000.
The experimental
backgrounds \cite {amen} for this reaction have been described; and their
minimization would lead to a high quality complementary determination of
F$_{3\pi}$.

The
$\gamma \pi \to \pi \pi^0$  reaction is also approved for study at CEBAF \cite
{rm} by measuring $\gamma p \to \pi^+ \pi^0 n$ cross sections near threshold
using tagged photons.
The accuracy of this method is limited by the
uncertainties associated with the needed Chew-Low \cite {cl} extrapolation to
the pion pole. F$_{3\pi}$ can also be studied at a low energy
electron-positron collider in the near threshold reaction $e^+ + e^-
\rightarrow \pi^+ + \pi^- +\pi^0$.

In conclusion, the experimental  result of Antipov et al. differs with the
chiral anomaly prediction  at O(p$^4$) by at least two standard deviations. At
O(p$^6$), considering the experimental and theoretical uncertainties, the data
and prediction are nearly consistent.
How well
does $\chi$PT work in the anomalous sector?
How anomalous is the real world anyhow?
We described how the Fermilab E781 experiment can give improved answers to
these questions via studies of reactions (4) and (7).
The 1 MHz pion
flux at Fermilab will enable significantly improved statistics, compared to
the previous experiments. In E781, at the 600 GeV
higher energy, and also at lower value of s; the strong contribution
to reaction (4) is
negligible \cite {in,ber}. This should significantly reduce the systematic
uncertainty.

\noindent
{\bf Acknowledgements} \\
\noindent
Discussions on this subject with J. Bijnens, L. Frankfurt,
S. Gerzon, B. Holstein, H. Leutwyler, S. Nussinov, E. Piasetzky, and J. Russ
are acknowledged. This work was supported by the U.S.-Israel Binational
Science Foundation (BSF), Jerusalem, Israel.

\end{document}